\documentclass[twocolumn,superscriptaddress,aps,showpacs,floatfix,pra,10pt,footnoteinbib]{revtex4-1}
\usepackage{amssymb,amsmath}
\usepackage{hyperref,graphicx}
\usepackage{dcolumn}
\usepackage{color}
\usepackage{ulem}
\usepackage{bbold}
\newcommand{\bs}[1]{{\boldsymbol{#1}}}
\newcommand{\bk}{\bs{k}}

\newcommand{\br}{\bs{r}}
\newcommand{\av}[1]{\overline{#1}}

\newcommand{\ket}[1]{\left| #1 \right\rangle}

\begin{document}

\title{Coherent backscattering reveals the Anderson transition}

\author{S. Ghosh}
\affiliation{Laboratoire Kastler Brossel, UPMC-Sorbonne Universit\'es, CNRS, 
ENS-PSL Research University, Coll\`{e}ge de France, 4 Place Jussieu, 75005 
Paris, France}
\affiliation{Centre for Quantum Technologies, National University of Singapore, 
3 Science Drive 2, Singapore 117543, Singapore}
\author{D. Delande}
\affiliation{Laboratoire Kastler Brossel, UPMC-Sorbonne Universit\'es, CNRS, 
ENS-PSL Research University, Coll\`{e}ge de France, 4 Place Jussieu, 75005 
Paris, France}
\author{C. Miniatura}
\affiliation{MajuLab, CNRS-UNS-NUS-NTU International Joint Research Unit, UMI 
3654, Singapore}
\affiliation{Centre for Quantum Technologies, National University of Singapore, 
3 Science Drive 2, Singapore 117543, Singapore}
\affiliation{Department of Physics, National University of Singapore, 2 Science 
Drive 3, Singapore 117542, Singapore}
\affiliation{INLN, Universit\'{e} de Nice-Sophia Antipolis, CNRS; 1361 route des 
Lucioles, 06560 Valbonne, France}
\author{N. Cherroret}
\email[Corresponding author: ]{cherroret@lkb.upmc.fr}
\affiliation{Laboratoire Kastler Brossel, UPMC-Sorbonne Universit\'es, CNRS, 
ENS-PSL Research University, Coll\`{e}ge de France, 4 Place Jussieu, 75005 
Paris, France}


\begin{abstract}
We develop an accurate finite-time scaling analysis of the angular width of 
the coherent backscattering (CBS) peak for waves propagating in 3D random media. 
Applying this method to ultracold atoms in optical speckle potentials, we show 
how to determine both the mobility edge and the critical exponent of the 
Anderson transition from the temporal behavior of the CBS width. Our 
method could be used in experiments to fully characterize the 3D Anderson 
transition.

\end{abstract}


\maketitle

In disordered media, the absence of diffusion arising from the spatial 
localization of single-particle states is known as Anderson localization (AL) 
\cite{Anderson58}. In three dimensions, AL manifests itself as a phase 
transition, which occurs at a critical energy, the mobility edge (ME), 
separating a metallic phase where  states are spatially extended, from an 
insulating one where states are localized. Theoretically, much efforts have been 
devoted to the study of the critical properties of the Anderson transition, 
such as wave functions at the ME \cite{Rodriguez09, Burmistrov11} or 
critical exponents \cite{Slevin14}. In practice however, only a handful of 
experiments have found evidence for the three-dimensional (3D) Anderson transition~\cite{Hu08, 
Aubry14, Jendrzejewski12, Kondov11, Semeghini14, Chabe08}. For matter waves, its 
critical (universal) features have been only investigated in the context of 
quantum-chaotic dynamical localization~\cite{Lemarie10}, but no such experiment 
in 3D disordered potentials has been reported to date.

In addition to the intrinsic difficulty of achieving wave localization in three 
dimensions, one reason for the rareness of experimental characterizations of the 
Anderson transition lies in the lack of easily measurable observables displaying 
criticality. In the context of atom optics, a routinely used approach consists 
in tracing the evolution in time of the spatial width of a spreading wave packet 
\cite{Jendrzejewski12, Kondov11, Semeghini14, Chabe08}. While AL implies a 
saturation of the width, the contrary is not true as classical effects can as 
well entail a saturation or a slowing down in time \cite{Muller14}. Furthermore, 
atomic wave packets have rather large energy distributions even when cooled down 
to very low temperatures, which forbids an accurate resolution of the critical 
region around the ME. Thus, any exploration of the Anderson transition 
with cold atoms should ideally be complemented with a clear demonstration of 
phase coherence, and should achieve a good energy resolution. For the latter 
issue, a first step has been reached in recent measurements of the ME 
based on a frequency modulation of the disorder \cite{Semeghini14} --
although the experimentally measured ME seems significantly higher than the one 
predicted from extensive numerical calculations~\cite{Delande14,Pilati15} -- 
and upcoming 
experiments are moving toward a genuine filtering of the energy distribution, 
required to access the critical properties of the transition~\cite{Josse15}. 
For additionally proving genuine phase-coherent scattering, the coherent 
backscattering (CBS) effect is a promising tool. CBS has already been observed 
in several experiments with cold atoms \cite{Jendrzejewski12_CBS}, light 
\cite{Albada85, Labeyrie99}, acoustic \cite{Bayer93} or seismic waves 
\cite{Larose04} in the (metallic) regime of diffusive transport. Interestingly 
however, CBS shows up not only in the metallic phase, but all the way across the 
Anderson transition. The question then naturally arises whether the CBS peak 
itself could be used as an observable for accessing the critical properties of 
this transition, in which case one would simultaneously ensure phase coherence. 
 \begin{figure}[h]
\begin{center}
\includegraphics[scale=0.25]{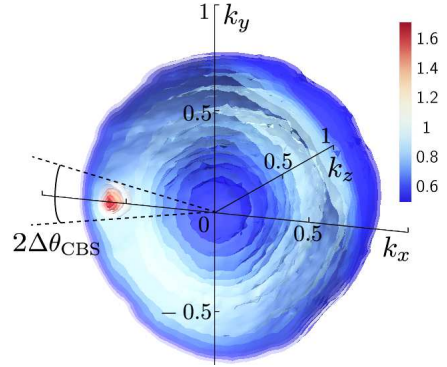}
\caption{(color online) Contour plot of the averaged momentum distribution of a 
matter wave, obtained after propagation of a plane wave $|\bk_0\rangle$ 
($\bk_0=0.6\hat{\boldsymbol{e}}_x$) in a speckle potential of strength $V_0=1$ 
for a duration $t=800$. The propagated state is here filtered around energy 
$E=-0.4$ (metallic regime). The CBS peak, of angular width 
$2\Delta\theta_\text{CBS}$, is visible at $\bk=-\bk_0$. Here momenta, energies 
and times are respectively in units of $\zeta^{-1}$, $\hbar^2/(m\zeta^2)$ and 
$m\zeta^2/\hbar$, where $\zeta$ is the correlation length of the potential.
}
\label{mom3d}
\end{center}
\end{figure}

In continuous-wave optical experiments, it is known that the CBS lineshape 
changes at the critical point \cite{Tiggelen00}. Unfortunately, this feature is 
usually smoothed by absorption or finite-size effects and cannot be used in 
practice. In this letter, we explore the dynamics of the CBS effect \emph{in momentum 
space} -- in  contrast with usual setups that search in configuration space -- around the
Anderson transition. By scrutinizing the dynamics of the CBS angular width, 
$\Delta\theta_\text{CBS}$, in combination with a numerical filter that provides 
a high energy resolution, we demonstrate that $\Delta\theta_\text{CBS}$ can 
be used to characterize the critical properties of the Anderson transition. By 
developing an accurate finite-time scaling analysis of the CBS data, we 
verify the one-parameter scaling theory of localization \cite{Abrahams79}, 
locate precisely the ME and extract the critical exponent of the 
transition. We determine these parameters for a speckle potential, and find a 
good agreement with the predictions of the transfer-matrix method.

As shown in \cite{Cherroret12}, CBS of cold atoms can be observed by tracing 
the evolution a quasi-plane matter wave in momentum space, a proposal recently 
realized experimentally \cite{Jendrzejewski12_CBS}. Let us thus consider a 
matter wave initially prepared in the plane-wave state 
$|\psi(t=0)\rangle=|\bk_0\rangle$, and subjected to a 3D random potential 
$V(\br)$. Following experiments, we choose $V(\br)$ to have the statistical 
properties of a blue-detuned speckle pattern. It is customary to shift all 
energies by the average value $V_0>0$ of the speckle potential, leading to the 
on-site distribution $P(V)=\exp[-(V+V_0)/V_0]\Theta(V+V_0)/V_0$ ($\Theta$ 
is the Heaviside function), and the two-point correlation function 
$\overline{V(\br)V(\br')} = V_0^2 
[\sin(|\br-\br'|/\zeta)/(|\br-\br'|/\zeta)]^2$, where $\zeta$ is the correlation 
length. 
In order to accurately pinpoint the ME $E_c$, it is useful to restrict
the evolution to a narrow energy range ($\pm\sigma$) centered at a given value 
$E$ that we wish to tune around the $E_c$, by applying a Gaussian filter~\cite{Ghosh14} $\exp[-(\hat{H}-E)^2/(2\sigma^2)]$
(where $\hat{H}=\hat{\textbf{p}}^2/(2m)+V(\textbf{r})$) on the initial state $|\bk_0\rangle.$
This filter makes it possible
to accurately extract $E_c$ which otherwise would be 
smoothed by the natural energy distribution of the initial plane wave in 
presence of the disordered potential \cite{Semeghini14, Skipetrov08}. 
Throughout this Letter, lengths, momenta, energies and 
times are given in units of $\zeta$, $\zeta^{-1}$, $\hbar^2/(m\zeta^2)$ and 
$m\zeta^2/\hbar$, respectively. We discretize the Hamiltonian $\hat{H}$ on a 3D 
grid of total volume $(60\times\pi\zeta)^3$ with periodic boundary conditions. 
Each cell of size $\pi \zeta$ is divided into $2$ steps in all three directions. 
In the following, we use $V_0=1,\sigma=0.02,k_0=0.6$.

The temporal evolution and the filtering are performed using a Chebyshev scheme.
The evolution operator over $\Delta t,$ $\mathrm{e}^{-i\hat{H}\Delta t/\hbar}$
[resp. the filtering operator] can be expanded in a series of Chebyshev polynomials of the first kind of 
$a\hat{H}\!+\!b$ [resp. $a(\hat{H}\!-\!E)^2\!+\!b$] with $a,b$ conveniently chosen parameters -- see~\cite{Roche97,Fehske09} for details --
whose coefficients are Bessel [resp. modified Bessel] functions of argument proportional to $\Delta t.$ 
The temporal evolution can be computed by iterating small time steps, each involving a limited number
of terms in the Chebyshev expansion.
The momentum wavefunction is obtained by Fourier transforming the final wave function $\ket{\psi(t)}$. The procedure is repeated
over $6\times10^3$  configurations of $V(\br)$, yielding the averaged momentum distribution  
$\av{n}(\bk,t)=\av{| \langle \bk \ket{\psi(t)}|^2}$. We show in Fig. \ref{mom3d} 
the numerical distribution $\av{n}(\bk,t)$ obtained at long  times for an energy 
$E=-0.4$ which lies in the metallic regime $E>E_c$. $\av{n}(\bk,t)$ clearly 
displays a narrow interference peak of angular width $\Delta\theta_\text{CBS}$ 
and centered at $\bk=-\bk_0$ (in red in Fig. \ref{mom3d}). This CBS peak sits on 
the top of a time-independent isotropic background (in blue in Fig. 
\ref{mom3d}), which in three dimensions has the shape of a spherical shell as a 
result of elastic multiple scattering off the random potential \cite{Cherroret12}.
\begin{figure}[h]
\begin{center}
\includegraphics[scale=.45]{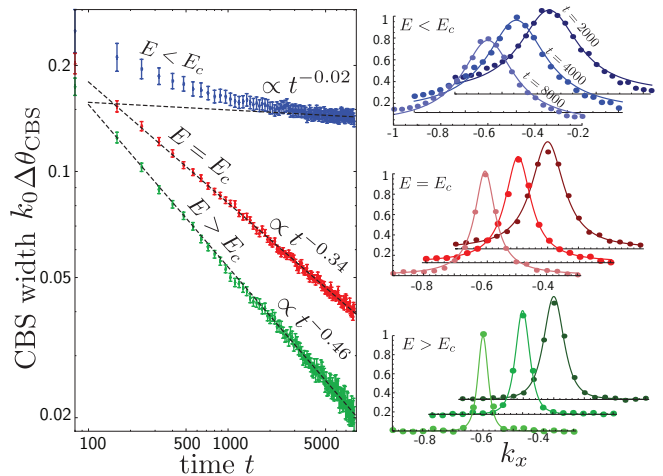}
\caption{(color online) Dynamics of the CBS peak across the Anderson transition. 
Left panel: angular width $\Delta\theta_\text{CBS}$ versus time, in the metallic 
regime $E=-0.4 > E_c$ (green points), at the mobility edge $E=E_c\simeq -0.48$ (red 
points), and in the insulating regime $E=-0.56<E_c$ (blue points). Right panels: 
cut along $k_x$ of the normalized CBS profile at three different energies. For 
each energy, profiles at three different times  $t=2000$, $4000$ and $8000$ are 
displayed, shifted with respect to each other for clarity. The CBS width rapidly 
saturates in the insulating regime, while it shrinks in time in the metallic and 
critical regimes. We find an excellent agreement with the temporal dependences 
predicted by Eq. (\ref{Delta_CBS}).}
\label{cbs_width}
\end{center}
\end{figure}

We now study the time dependence of the CBS angular width, 
$\Delta\theta_\text{CBS}$. Qualitatively, CBS is an interference effect between 
two waves that propagate along an identical multiple scattering sequence 
$\br_1\ldots\br_N$ but in opposite directions \cite{John91}. The interference 
term between these paths is proportional to 
$\av{\cos[(\bk_0+\bk)\cdot(\br_N-\br_1)]}$. Therefore, denoting by 
$\Delta\theta$ the angle (assumed small) between $\bk$ and $-\bk_0$, we infer 
that an interference is visible on average provided $k_0\Delta\theta\Delta r(t)\ll1$, 
where $\Delta r(t)=(\av{|\br_N(t)-\br_1|^2})^{1/2}$. We thus estimate the 
angular width of the CBS at a given time $t$ to be 
$\Delta\theta_\text{CBS}\sim1/[k_0\Delta r(t)]$. The average distance between 
the first and last points of the scattering sequence depends on the nature of 
transport in the system. In the metallic regime $E>E_c$, $\Delta r(t)\propto 
\sqrt{D(E)t}$ with $D(E)$ the diffusion coefficient at energy $E$, while  
$\Delta r(t)\propto t^{1/3}$ at $E=E_c$ \cite{Ohtsuki97} and 
$\Delta r(t)\propto\xi(E)$, the localization length, in the insulating regime 
$E<E_c$. We thus have:
\begin{equation}
\label{Delta_CBS}
k_0\Delta\theta_\text{CBS} \sim
\begin{cases} 
1/\sqrt{D(E) t} & E>E_c \\
 1/t^{1/3} &  E=E_c \\
1/\xi(E) & E<E_c.
\end{cases}
\end{equation}
The time dependence of $\Delta\theta_\text{CBS}$ is thus qualitatively different 
in the three regimes of transport. In particular, a sub-diffusive behavior of 
the CBS width marks the position of the ME $E_c$. We have performed 
numerical simulations of the momentum distribution for various energies $E$ 
around $E_c\simeq-0.48$.
We show in the left panel of Fig. \ref{cbs_width} the CBS width as a function of 
time, for three different energies around $E_c$. At long times, the results 
follow very well the predictions of Eq. (\ref{Delta_CBS}). For each energy, we 
have obtained $\Delta\theta_\text{CBS}$ by first removing the isotropic 
background~\cite{Ghosh14} from the 3D momentum distribution, then fitting the 
resulting momentum profile with $\alpha/[1+(\bk+\bk_0)^2/\beta]^\gamma$ (where 
$\alpha$, $\beta$ and $\gamma$ are time- and energy-dependent fit parameters), 
and finally taking the half width at half maximum of the fitting function. 
Error bars on $\Delta\theta_\text{CBS}$ have been estimated from the standard 
deviations of $\beta$ and $\gamma$. We show examples of CBS profiles and the 
corresponding fits in the right panel of Fig. \ref{cbs_width}.

\begin{figure}[h]
\begin{center}
\includegraphics[scale=.47]{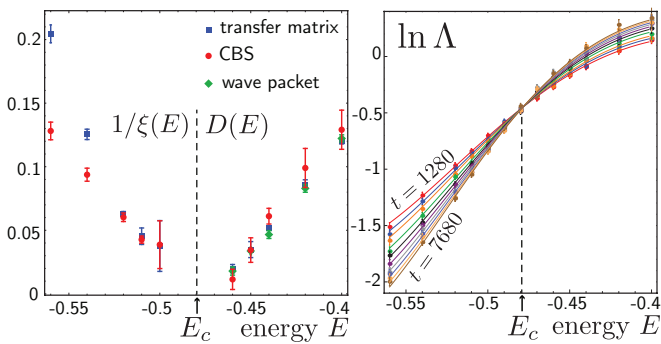}
\caption{(color online) Left panel: diffusion coefficient (in the metallic 
regime $E>E_c$), and inverse of the localization length $1/\xi(E)$ (in the 
insulating regime $E<E_c$) versus energy $E$. Red dots are obtained from an 
analysis of the CBS width, blue squares from the transfer-matrix method, and 
green diamonds from the spatial spreading of a wave packet. Right panel: scaling 
function $\Lambda$ versus energy around the ME, for various times ranging 
from $t=1280$ to $t=7680$. The curves cross at a common point $E_c \approx 
-0.48$, which signals the location of the ME. 
Points are the results of numerical simulations of CBS, while solid curves are 
fits of these data using Eq.~(\ref{scaling_function}).
\label{Fig_3}}
\end{center}
\end{figure}
According to Eq. (\ref{Delta_CBS}), the CBS width is also proportional to the 
square root of the diffusion coefficient $D(E)$ in the metallic regime, and to 
the inverse of the localization length $1/\xi(E)$ in the insulating regime, 
which suggests an original way of measuring these quantities experimentally. To 
demonstrate the efficiency of such an approach, we have extracted $D(E)$ and 
$1/\xi(E)$ from the numerical data for $\Delta\theta_\text{CBS}$, by 
extrapolating the quantities $1/[(k_0\Delta\theta_\text{CBS})^2t]$ (for $E>E_c$) 
and $1/(k_0\Delta\theta_\text{CBS})$ (for $E<E_c$) to infinite times. The 
results are shown as red dots in the left panel of Fig. \ref{Fig_3}, for various 
energies around $E_c$ (no values too close to $E_c$ are shown due to the lack of 
accuracy of the extrapolation procedure at these energies. The vicinity of $E_c$ 
deserves a special analysis that will be described below). We have computed 
these quantities using the transfer-matrix method  (blue squares) \cite{Mkk83}. 
In the metallic region, we have also computed $D(E)$ by yet another method that 
consists in analyzing the spatial width of a spreading, initially narrow wave 
packet as a function of time (green diamonds). All the results for $D(E)$ are in 
very good agreement. The predictions below $E_c$ tend to deviate far from the 
ME, which we explain by the difference in the definition of $\xi(E)$ in 
the two methods: the localization length that appears in 
$\Delta\theta_\text{CBS}$ controls the exponential decay of the average density, 
whereas the localization length that appears in transfer matrices controls the 
exponential decay of the average of the logarithm of the transmission 
\cite{Mkk83}.

Let us now explore the behavior of $\Delta\theta_\text{CBS}$ in the close 
vicinity of $E_c$. In this region, $D(E)\propto |E-E_c|^s$ and 
$\xi(E)\propto|E-E_c|^{-\nu}$, where the two critical exponents $\nu$ and $s$ 
turn out to be equal for the Anderson transition in dimension 3 \cite{Wegner76}. 
Near $E_c$, the three scaling laws (\ref{Delta_CBS}) can be recast under 
the unified form
\begin{eqnarray}
\Lambda\equiv \frac{1}{Lk_0\Delta\theta_\text{CBS}}= 
F\left[\chi_r(E)L^{\, 1/\nu} \right],
\label{scaling_law}
\end{eqnarray}
where $\chi_r(E)\propto E-E_c$, $L=[t/(2\pi\hbar\rho(E))]^{1/3}$ with $\rho(E)$ 
the density of states per unit volume at energy $E$, and $F$ is a function 
characteristic of the transition. Although the system a priori depends on two 
parameters $E$ and $t$, Eq. (\ref{scaling_law}) thus suggests that  $\Lambda$ is 
in fact a function of a {\it{single}} parameter, and is therefore a good 
candidate for developing a single-parameter scaling description of the Anderson 
transition \cite{Abrahams79}. The introduction of the length scale $L$ 
\cite{footnote2} allows us to establish a straightforward analogy with the usual 
scaling theory of Anderson localization for time-independent disordered systems 
\cite{Abrahams79, Slevin14}. A direct consequence of Eq. (\ref{scaling_law}) is  that when $\ln\Lambda$ is 
plotted against $E$, the curves at different times should cross at $E=E_c$. This 
behavior is well visible in the right panel of Fig. \ref{Fig_3}. By pinpointing 
the location of the crossing, we obtain a first estimation of the ME: $E_c\simeq-0.48$. 

Guided by the one-parameter scaling theory of Anderson localization 
\cite{Abrahams79}, we now postulate that Eq. (\ref{scaling_law}) holds not only 
in the close vicinity of the ME [where $\chi_r(E)\propto E-E_c$] but also 
away from it, and propose to verify this hypothesis by a rigorous finite-size 
scaling analysis of the numerical data for the CBS width. 
For this purpose, we introduce a fitting function of the data by Taylor 
expanding Eq. (\ref{scaling_law}) up to a certain order $n_R$~\cite{Slevin14},
\begin{eqnarray}
\Lambda= \sum_{n=0}^{n_R}\chi_r(E)^{n} L^{n/\nu} F_{n},
\label{scaling_function}
\end{eqnarray}
and further expand the variable $\chi_r(E)$ according to 
$\chi_{r}(E)=\sum_{m=1}^{m_R} b_{m} (E_c-E)^m$.
In this model, $F_n$, $b_m$, $\nu$ and $E_c$ are free parameters. We determine 
them using a least-square fit of the data for $\Lambda$ with Eq. 
(\ref{scaling_function}) retaining data only for sufficiently long times (such 
that $L>20$). We show in the right panel of Fig. \ref{Fig_3} the results of this 
fit for curves $\ln\Lambda$ versus $E$ (solid lines). We used 
$n_R\!=\!2,m_R\!=\!3$ (that is 7 fitting parameters) for 1141 data points. The 
$\chi^2$ per degree of freedom
is found to be 0.55. This small value (from the statistical significance 
point of view) comes from the fact that the data got at the same energy, but 
different sizes (i.e. different times), are obtained
using the same realizations of the disordered potential and thus have residual 
correlations. We have also tried to include irrelevant scaling variables to 
better account for deviations to scaling expected at short times  
\cite{Slevin14, Lemarie09}, but we did not observe significant improvements of 
the quality of the fits.

\begin{figure}[h]
\begin{center}
\includegraphics[scale=.40]{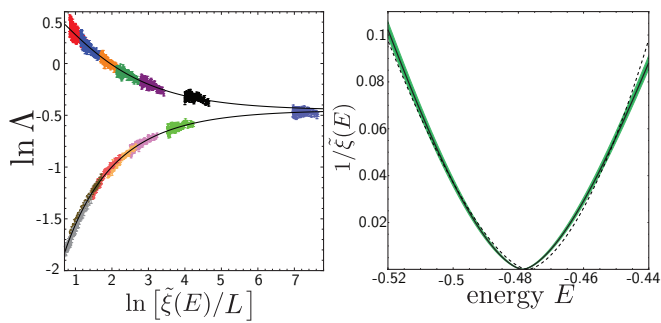}
\caption{(color online) Left panel: scaling function $\Lambda$ constructed by 
fitting the data for $\Lambda$ with model (\ref{scaling_function}). Points are 
the data, and the solid black curve is the fit. All data lie on the same master 
curve, in agreement with the one-parameter scaling hypothesis, Eq. 
(\ref{scaling_law}). Right panel: $1/\tilde\xi(E)=|\chi_r(E)|^{\nu}$ versus 
energy $E$ (solid curve), together with the confidence interval (shadowed 
region, green online). $1/\tilde\xi(E)$ vanishes at the ME, and is 
proportional to $|E-E_c|^{\nu}$ in its vicinity. 
The dashed curve is the prediction obtained from an independent finite-size 
scaling analysis based on the transfer-matrix approach.}
\label{ScalingF_xi}
\end{center}
\end{figure}

We then plot the data $\ln\Lambda$ as a function of $\tilde\xi(E)/L$, where 
$\tilde\xi(E)=|\chi_r(E)|^{-\nu}$ (colored points), together with the fit to 
model (\ref{scaling_function}) (solid curve). The results are shown in the left 
panel of Fig. \ref{ScalingF_xi}. We see that all data collapse almost perfectly 
on the same master curve. This result demonstrates that the function $\Lambda$, 
as computed from the width of the CBS peak, does follow the one-parameter 
scaling theory, in full agreement with Eq. (\ref{scaling_law}). The quantity 
$\tilde\xi(E)$ is proportional to the localization length $\xi(E)$ on the 
insulating side of the transition, and proportional to the inverse of the 
diffusion coefficient, $1/D(E)$, on the metallic side. In the right panel of 
Fig. \ref{ScalingF_xi} we show $1/\tilde\xi(E)$ as a function of energy, as 
obtained from the fitting procedure. As expected, $1/\tilde\xi(E)$ vanishes at 
$E=E_c$, which signals the divergence of the localization length and the 
vanishing of the diffusion coefficient. 
The fitting analysis also allows us to provide estimations of $E_c$ and  of the critical exponent $\nu$. We find $E_c=-0.4786\pm 13.10^{-4}$ and $\nu=1.61\pm0.03$. Because the above-mentioned chi-squares are too small, they cannot 
be used to extract the uncertainty. We have thus divided the whole configuration sample in several independent subsets,
and estimated $E_c$ and $\nu$ for each subset. The reported uncertainties reflect
the deviations between the different subsets. They are found to weakly depend on $\sigma$, most
probably because
the finite size scaling approach relies on data belonging to an energy interval 
much larger than $\sigma$. In the right panel of Fig.~\ref{ScalingF_xi}, we 
also display as a dashed curve the quantity $\tilde\xi(E)$ computed from an 
independent finite-size scaling analysis based on the transfer-matrix method 
\cite{Slevin14, Lemarie09}. The latter provides $E_c=-0.4771 \pm 7.10^{-4}$ and 
$\nu=1.62 \pm 0.03$, in a somewhat surprisingly good agreement with the estimations extracted from the 
CBS width. The slight discrepancy from the recently reported value 
$E_c=-0.43$~\cite{Delande14} comes from the crude discretization we used to save 
computer resources. Indeed, as involving a time propagation and a narrow energy 
filter, the characterization of the Anderson transition from the CBS peak is 
more numerically demanding than from the transfer-matrix approach. Due to this 
discretization the free-space dispersion relation deviates from the massive one 
$E=k^2/2$ and $\rho(E)$ is overestimated near the ME, lowering $E_c$. This 
shift has however no effect on the physics of the CBS effect or on the Anderson 
transition.

In conclusion, we have shown that the dynamics of the CBS peak can be used to 
characterize the Anderson transition, enabling to (i) accurately pinpoint the 
location of the ME (ii) access the critical exponent and (iii) test the 
validity of the single-parameter scaling hypothesis. Our method has the double 
advantage to be based on a physical observable --the CBS peak-- which is usually 
well controlled in experiments, and to demonstrate phase coherence, which is a 
crucial requirement prior any claim for Anderson localization. The approach has 
straightforward applications to the field of atom optics in disordered 
potentials, but it can also be applied to the context of localization of 
classical waves \cite{Hu08}.

The authors thank C.A. M\"uller and V. Josse for helpful discussions. SG 
acknowledges the support of the PHC Merlion Programme of the French Embassy in 
Singapore. This work was granted access to the HPC resources of TGCC under the 
allocation 2015-057083 made by GENCI (Grand Equipement National de Calcul 
Intensif) and to the HPC resources of MesoPSL financed by the Region Ile de 
France and the project Equip@Meso (reference
ANR-10-EQPX-29-01) of the programme Investissements d'Avenir supervised by the 
Agence Nationale pour la Recherche. The Centre for Quantum Technologies is a 
Research Centre of Excellence founded by the Ministry of Education and the 
National Research Foundation.


\begin{thebibliography}{99}
\bibitem{footnote2}
$L(t)$ is nothing but the size of a system for which the Heisenberg time is 
equal to $t$.
\end{thebibliography}
\end{document}